\documentclass[twocolumn,prl,preprintnumbers,tightenlines,nofootinbib,superscriptaddress]{revtex4-2}

\usepackage[colorlinks=true,breaklinks=true]{hyperref}
\hypersetup{allcolors=[rgb]{0.0 0.0 1.0},linkcolor=[rgb]{0.75 0.05 0.05}}
\usepackage{orcidlink}
\usepackage{graphicx,wrapfig,float,slashed,subcaption,bbold,bm}
\usepackage{amsmath,amssymb,epsfig,graphicx,xcolor}
\usepackage{epstopdf}
\usepackage{booktabs}
\usepackage{ragged2e}
\usepackage{mciteplus}
\usepackage{mathtools}
\usepackage{enumitem}
\usepackage{cleveref}

\usepackage{cancel}

\usepackage{hyperref} 
\usepackage{nameref}

\hypersetup{
    colorlinks=true,
    allcolors=blue
}

\usepackage{makecell}
\usepackage{multirow}
\usepackage{adjustbox}

\usepackage{comment}

\newcommand{\nc}{\newcommand}
\nc{\non}{\nonumber}
\nc{\hc}{\hbox {H.c.}}
\nc{\noi}{\noindent}
\nc{\barx}{\bar{x}}
\nc{\pbarn}{\;\hbox {pb}}
\nc{\fbarn}{\;\hbox {fb}}

\nc{\hsp}{\hspace{0.5cm}}
\nc{\lsp}{\hspace{1cm}}
\nc{\Lsp}{\hspace{2cm}}
\nc{\LLsp}{\lsp\lsp}
\nc{\lra}{\longrightarrow}
\nc{\p}{\prime}
\nc{\sgn}{\text{sgn}}
\nc{\ph}{\varphi}
\nc{\op}{{\cal O}}

\nc{\beq}{\begin{equation}}  \nc{\eeq}{\end{equation}}
\nc{\bea}{\begin{eqnarray}}  \nc{\eea}{\end{eqnarray}}
\nc{\baa}{\begin{array}}     \nc{\eaa}{\end{array}}
\nc{\bit}{\begin{itemize}}   \nc{\eit}{\end{itemize}}
\nc{\ben}{\begin{enumerate}} \nc{\een}{\end{enumerate}}
\nc{\bce}{\begin{center}}    \nc{\ece}{\end{center}}
\nc{\bpm}{\begin{pmatrix}}   \nc{\epm}{\end{pmatrix}}
\nc{\bvt}{\begin{verbatim}}  \nc{\evt}{\end{verbatim}}
\nc{\bealg}{\begin{equation}\begin{aligned}} 
\nc{\eealg}{\end{aligned}\end{equation}} 

\definecolor{darkgreen}{rgb}{0,0.5,0}

\def\lsim{\mathrel{\raise.3ex\hbox{$<$\kern-.75em\lower1ex\hbox{$\sim$}}}}
\def\gsim{\mathrel{\raise.3ex\hbox{$>$\kern-.75em\lower1ex\hbox{$\sim$}}}}

\def\udots{\mathinner{\mkern1mu\raise1pt\vbox{\kern7pt\hbox{.}}\mkern2mu\raise4pt\hbox{.}\mkern2mu\raise7pt\hbox{.}\mkern1mu}}

\def\ev{\;\hbox{eV}}

\def\mev{\;\hbox{MeV}}
\def\gev{\;\hbox{GeV}}

\def\dd{\mathrm d}

\newcommand{\Eqs}[2]{Eqs.~(\ref{#1}) and (\ref{#2})}

\newcommand{\UV}{\mathrm{UV}}
\newcommand{\IR}{\mathrm{IR}}
\newcommand{\PQ}{\mathrm{PQ}}
\newcommand{\QCD}{\mathrm{QCD}}
\newcommand{\YM}{\mathrm{YM}}

\global\long\def\b#1{\left(#1\right)}%

\global\long\def\pd{\partial}

\begin{document}

\title{The Holographic QCD Axion in Five Dimensions}

\author{Csaba Cs\'aki\orcidlink{0000-0001-8899-6073}}
\email{csaki@cornell.edu}
\affiliation{Laboratory for Elementary Particle Physics, Cornell University, Ithaca, NY 14853, USA}
\author{Eric Kuflik\orcidlink{0000-0003-0455-0467}}
\email{eric.kuflik@mail.huji.ac.il}
 \affiliation{Racah Institute of Physics, Hebrew University of Jerusalem, Jerusalem 91904, Israel}
\author{Wei Xue\orcidlink{0000-0003-1568-4946}} 
\email{weixue@ufl.edu}
\affiliation{Institute for Fundamental Theory, 
University of Florida, Gainesville, FL 32611, USA}
\author{Taewook Youn\orcidlink{0000-0003-2229-5025}}
\email{taewook.youn@cornell.edu}
\affiliation{Laboratory for Elementary Particle Physics, Cornell University, Ithaca, NY 14853, USA}
\affiliation{School of Physics, Korea Institute for Advanced Study, Seoul 02455, Republic of Korea}

\begin{abstract}
We present a holographic construction of the QCD axion based on a warped 5D model. A key ingredient of our setup is the introduction of a bulk scalar field $\theta$, which is holographically dual to the topological operator of QCD. This makes the relation among the axion, the $\eta'$, and the anomalies transparent. We identify the bulk modes corresponding to the $\eta'$ and axion states, and show that an adjustment analogous to that of the usual 4D axion takes place. We identify the origin of the axion quality problem in this framework and show that a large degree of axion compositeness is needed to solve it. We also find that, in the limit of a high quality axion, the physical axion state is predominantly contained in the bulk gauge field. 

\end{abstract}


\maketitle

\section{Introduction}

Axions might be the key to unlocking some of the deepest mysteries plaguing the Standard Model (SM) of particle physics. They provide the most elegant solution to the strong CP problem~\cite{Peccei:1977hh,Peccei:1977ur,Wilczek:1977pj,Weinberg:1977ma}, and could at the same time play the role of dark matter (DM). The vacua of quantum chromodynamics (QCD) are characterized by a parameter $\theta$, which along with the coefficient of the ${\rm Tr}\,G\widetilde{G}$ operator in the Lagrangian and the phase of the fermion mass matrix determinant combine to a physically observable $\bar{\theta}$ parameter, that will contribute to the measurement of the CP violating neutron electric dipole moment. The null observation of a neutron electric dipole moment implies $\bar{\theta} \lsim 10^{-10}$~\cite{Abel:2020pzs}, leading to the strong CP problem: while $\bar{\theta}$ could a priori take on any value, measurements restrict it to be extremely tiny without a clear symmetry explanation (since CP is broken in other sectors of the SM).

The presence of the axion - an extremely light and very weakly coupled particle that is the Goldstone boson of a global U(1)$_{\PQ}$ symmetry which is anomalous under the $SU(3)_{\QCD}$ strong interactions -  would beautifully remedy this issue~\cite{Kim:1979if,Shifman:1979if,Dine:1981rt,Zhitnitsky:1980tq}. The axion would naturally settle to its minimum where the effective value of $\bar{\theta}$ naturally vanishes.  This same axion could also constitute all of the observed DM in the universe~\cite{Preskill:1982cy,Abbott:1982af,Dine:1982ah}, and due to its very weak coupling is poorly constrained by traditional collider physics experiments. Instead dedicated axion searches have to be developed, which has led to one of the major programs of modern particle physics (for recent reviews, see~\cite{Ringwald:2024uds,Giannotti:2024xhx,Berlin:2024pzi,Caputo:2024oqc,Dobrich:2025oso}). In spite of these major successes the axion paradigm faces a serious theoretical challenge: since its coupling is extremely weak, the U(1) symmetry protecting the axion has to be close to exact; even tiny explicit breaking terms will spoil the axion sliding to the correct minimum. This is usually called the axion quality problem~\cite{Georgi:1981pu,Kamionkowski:1992mf,Barr:1992qq,Holman:1992us}, and one usually needs to find a separate mechanism to solve the quality problem. The most promising such solutions are to make the axion composite, or to consider an axion that arises from a component of a gauge field, usually denoted as $A_5$ upon compactification of an additional spatial dimension~\cite{Kim:1984pt,Choi:1985cb,Randall:1992ut,Choi:2003wr,Redi:2016esr,Lillard:2017cwx,Gavela:2018paw,Cox:2019rro,Bigazzi:2019eks,Ardu:2020qmo,Yin:2020dfn,Gherghetta:2020ofz,Contino:2021ayn,Lee:2021slp,Gherghetta:2025fip,Gherghetta:2025kff,Agrawal:2025mke}. Such $A_5$'s are ubiquitous in string theory, and are often also referred to as ``string theory axions"~\cite{Svrcek:2006yi,Arvanitaki:2009fg,Demirtas:2021gsq}. 

A substantial literature has explored axions in extra-dimensional theories, beginning already in the era of large extra dimensions~\cite{Dienes:1999gw}. Representative constructions include flat bulk axions, extra-dimensional realizations with a gauged PQ symmetry, axions arising from higher-dimensional gauge fields or from bulk scalar sectors, in both flat and warped backgrounds~\cite{Cheng:2001ys,Flacke:2006ad,Bigazzi:2019eks,Cox:2019rro,Bonnefoy:2020llz,Lee:2021slp,deGiorgi:2024elx}. 
While a top-down string theory-based holographic QCD axion has been studied in~\cite{Bigazzi:2019eks},
surprisingly the most basic 5D holographic implementation of the QCD axion using the 5D AdS/QCD correspondence~\cite{Erlich:2005qh,DaRold:2005mxj} has so far not been considered.  The aim of this paper is to present this simple construction of the 5D holographic QCD axion, and to examine the quality problem in this context. We will find, similar to the results of~\cite{Cox:2019rro}, that a high quality axion will require a large degree of compositeness, which in the holographic theory will also imply that a high quality axion is mostly consisting of a ``stringy" $A_5$ axion. 

The key ingredient of our 5D model is the representation of the axial and PQ anomalies in the holographic theory. The $\theta$-angle, the source of the ${\rm Tr}~G\widetilde{G}$ operator of QCD, will become a full-fledged (real) bulk scalar, while both the $U(1)_A$ and $U(1)_{\PQ}$ global symmetries will be bulk gauge symmetries. As we will explain, the anomaly will be represented by a St\"uckelberg term whereby the bulk $\theta$ field will pick up a shift symmetry under both anomalous U(1)'s.  It will be the interplay of this bulk shift symmetry with the UV boundary condition of the bulk $\theta$ field that will give rise to the holographic description of the axial anomaly.

\section{The 5D Action}
\begin{figure}[t!]
  \centering
  \includegraphics[width=0.8\columnwidth]{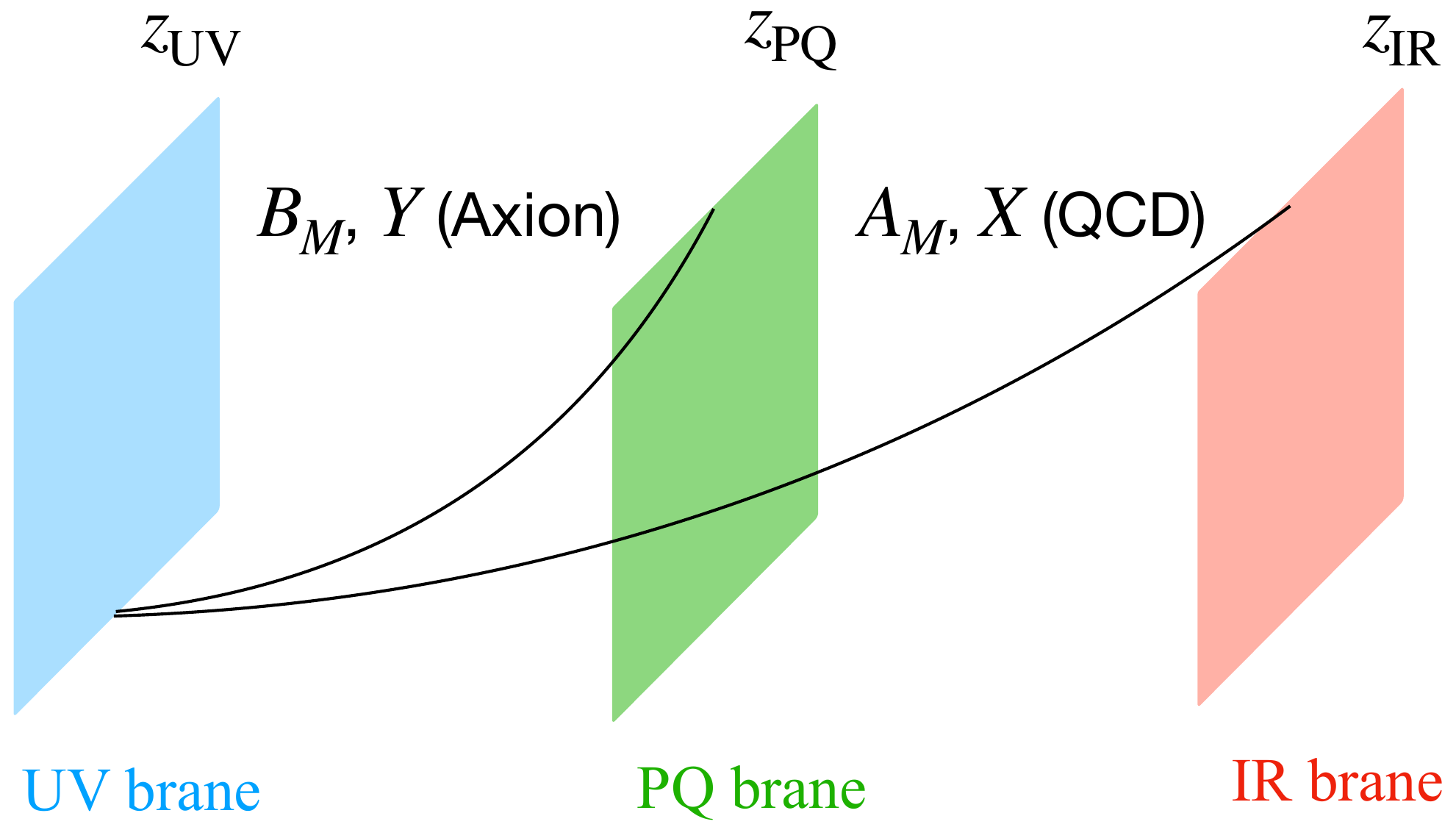}
  \caption{\protect\RaggedRight Illustration of the brane configuration: the IR brane sets the QCD scale, while the PQ brane sets the scale for the axion physics.}
  \label{fig:brane}
\end{figure}

We work with a five-dimensional $U(1)_A \times U(1)_{\PQ}$ gauge theory on a slice of AdS,
\beq
\dd s^2=\frac{R^2}{z^2}(\eta_{\mu\nu}\dd x^\mu \dd x^\nu- \dd z^2),
\eeq
where $R$ is the AdS curvature radius and $z \in [z_{\UV},z_{\IR}]$, as in the original Randall-Sundrum model~\cite{Randall:1999ee}. An intermediate PQ brane at $z = z_{\PQ}$ will localize the Peccei-Quinn sector.\footnote{For an early warped extra dimensional axion with an intermediate brane see~\cite{Collins:2002kp}.} The bulk contains two Abelian gauge fields---$A_M$, dual to the singlet axial current $J_5^\mu$, and $B_M$, dual to the $U(1)_{PQ}$ current. For an illustration of our setup see FIG.~\ref{fig:brane}. 
In addition, there are 3 scalar fields:
\begin{itemize}
  \item $\theta$, dual to the topological operator $G \tilde G$. The value of this scalar field on the UV brane will correspond to $\theta_{QCD}$.
  \item $X \equiv \frac{\rho_X}{\sqrt2} e^{i\phi_X}$, dual to the quark bilinear $\bar q q$. The background profile of $\rho_X$ encodes the spontaneous breaking of chiral symmetry via the quark condensate, while a source on the UV brane will correspond to the quark masses explicitly breaking the axial $U(1)_A$. 
  \item  $Y \equiv \frac{\rho_Y}{\sqrt2} e^{i \phi_Y}$, dual to the PQ order parameter. The background profile of $\rho_Y$ encodes the spontaneous breaking of the PQ symmetry.  The phase, $a$, will contain the 4D-axion, which mixes with the fifth components of the gauge fields $B_z,A_z$ as well as $\theta , \phi$. 
\end{itemize}

The Lagrangian is given by
\begin{widetext}
\begin{eqnarray}
 S_\mathrm{5D} &&=\int \dd^5 x \sqrt{|g|} \left[ \frac{g_\theta^2}{2}(\pd_M \theta - \kappa_A A_M - \kappa_B B_M )(\pd^M \theta - \kappa_A A^M  -  \kappa_B B^M)  - \frac{1}{4e_A^2} F^{MN}_A{F_{A}}_{MN}  - \frac{1}{4 e_B^2} F^{MN}_B {F_{B}}_{MN}
 \right. \nonumber\\
&&\left.  
+ \frac{\rho_X^2}{2} (\pd_M \phi_X - q_A A_M)(\pd^M \phi_X - q_A A^M)   + \frac{\rho_Y^2}{2} (\pd_M \phi_Y - q_B B_M)(\pd^M \phi_Y - q_B B^M) + \mathcal{L}_5(\rho_X, \rho_Y)
+ {\rm g.f.} 
\right],
\end{eqnarray}
\end{widetext}
where $\mathcal{L}_5(\rho_X, \rho_Y)$ contains kinetic and potential terms for the radial fields $\rho_X$ and $\rho_Y$. The gauge fixing terms, $ {\rm g.f.} $, is presented in Appendix \ref{app:RxiGauge}.
Solving their background equations of motion yields profiles characterized by a spontaneous symmetry-breaking vacuum expectation value (VEV) $v$ and an explicit symmetry-breaking UV source $s$. To leading order, these general solutions take the form
\begin{equation}
\rho(z) = \frac{R^{-\frac{3}{2}}}{\sqrt{2}} \left[ v \, z_{\rm IR, PQ} \left(\frac{z}{z_{\rm IR, PQ}}\right)^{\Delta} + s \, z_{\rm UV} \left(\frac{z}{z_{\rm UV}}\right)^{4-\Delta} \right]\,,
\end{equation}
where $\Delta$ is the conformal dimensions of the dual operators (for instance, $\Delta_X=3$ corresponds to the standard quark condensate).  

The coefficients $\kappa_{A,B}$ encode the chiral anomalies of the $U(1)_A$ and $U(1)_{PQ}$ symmetries under QCD (or its large $N$ version considered here). 
Under bulk $U(1)_{A,B}$ gauge transformations $A_M\to A_M +\partial_M \epsilon_A$ and $B_M\to B_M +\partial_M \epsilon_B$ the field $\theta$ will pick up a shift 
\begin{equation}
\theta \to \theta + \kappa_A \epsilon_A +\kappa_B \epsilon_B
\label{eq:shift}
\end{equation}
which will ensure that the entire bulk action remains gauge invariant. The anomaly of the symmetries manifests itself in the shift of the boundary value of $\theta$ under (\ref{eq:shift}): $\theta|_{\UV}\to \theta|_{\UV}+\kappa_A \epsilon_A|_{\UV}+\kappa_B \epsilon_B|_{\UV}$. Since $\theta|_{\UV}$ is the source for $G\tilde{G}$ in the CFT, this shifted boundary value implies the generation of the usual anomaly term on the CFT side. In this way, the effect of the anomaly in the bulk is incorporated as a simple St\"uckelberg term.\footnote{Note however, that one can consider the Hodge dual of the $\theta$ field, for which the St\"uckelberg term will turn into the more familiar topological Chern-Simons-like $BF$-term, see~\cite{Csaki:2026arv}.}

To encode the effect of explicit symmetry breaking, UV-brane localized potentials are added
\beq
S_{\rm UV} = \int \dd^4x\sqrt{|g^{\rm ind}|}\big[  - V(X) -U(Y)\big]\,,
\eeq
where $\sqrt{|g^{\rm ind}|}= R^4/z^4_\UV$, is the induced metric on the brane. The UV potential for the $X$ scalar will correspond to the quark masses explicitly breaking the $U(1)_A$ axial symmetry 
\beq
V(X) \supset -m_q R^{-\frac{3}{2}}X^\dagger + h.c.,
\eeq
while $U(Y)$ includes explicit PQ breaking terms, which are responsible for the PQ quality problem. 

To ensure a well-defined variational principle and systematically decouple the unphysical longitudinal gauge modes from the physical scalar fluctuations, we also include a generalized $R_\xi$ gauge-fixing term in the bulk action (the explicit form and its role in the zero-mode derivation are detailed in the Appendix).

To complete the description of the model one needs to specify the boundary conditions of the various bulk fields on the UV and IR branes, as well as the intermediate PQ brane. The $U(1)_A$ axial symmetry is a global symmetry, that is broken by the $X$ field, but otherwise unbroken. Hence its boundary condition on the UV brane should eliminate a potential gauge zero mode, requiring a Dirichlet boundary condition, while it should be unbroken (flat) on the IR brane. To satisfy the full variational principle in the presence of the $R_\xi$ gauge-fixing term, the variation with respect to $A_z$ on the UV brane yields an associated constraint, leading to
\bealg
A_\mu|_{\UV} = 0, \\
\left. -\frac{1}{e_A^2} \partial_z\left(\frac{R}{z} A_z\right) + \frac{R^3}{z^3} (\rho_X^2 q_A \phi_X + g_\theta^2 \kappa_A \theta) \right|_{\UV} = 0, \\
\partial_z A_\mu|_{\IR} = 0, \quad A_z|_{\IR} = 0\,.
\eealg
The gauge field for the $U(1)_{\PQ}$ global symmetry obeys identical boundary conditions, except that its geometry terminates on the intermediate PQ brane rather than the IR brane:
\bealg
B_\mu|_{\UV} = 0, \\
\left. -\frac{1}{e_B^2} \partial_z\left(\frac{R}{z} B_z\right) + \frac{R^3}{z^3} (\rho_Y^2 q_B \phi_Y + g_\theta^2 \kappa_B \theta) \right|_{\UV} = 0, \\
\partial_z B_\mu|_{\PQ} = 0, \quad B_z|_{\PQ} = 0\,.
\eealg

As discussed above, the value of the bulk $\theta$ scalar on the UV brane is identified with the source for the CP violating operator ${\rm Tr}\, G\tilde{G}$ in the CFT
\beq
\theta|_{\UV}  = \theta_{\rm QCD}\,.
\eeq
The most important and non-trivial BC for $\theta$ is the one on the IR brane. This boundary can not be easily guessed - one rather needs to rely on the string theory implementation~\cite{Witten:1998uka,Witten:1998zw,Bartolini:2016dbk} of holographic models of QCD. There, instead of a single extra dimension there is whole disc, and the location of the IR brane here corresponds to the center of the disc. The bulk $\theta$ is a Wilson loop around on this disc, which shrinks to zero at the IR brane. Hence for pure Yang-Mills one is led to the BC $\theta|_{\IR}=0$.\footnote{For a more complete discussion see~\cite{Csaki:2026arv, Mishra:2026lvq}.} In the presence of the scalars $X$ and $Y$ $\theta$ picks up a non-trivial shift under the chiral U(1) symmetries, and the BC  apply to the gauge invariant combination 
\beq
\Omega(z) \equiv \theta(z) - \frac{\kappa_A}{q_A}\phi_X(z) - \frac{\kappa_B}{q_B}\phi_Y(z)\,.
\label{Omega}
\eeq
Since the $a$ field terminates on the PQ brane, the final form of the IR BC will be 
\beq
\left.\left(\theta - \frac{\kappa_A}{q_A}\phi_X\right)\right|_{\IR} = 0\,.
\eeq

The BC's of the angular fields $\phi$ and $a$ on the UV brane are determined by the localized explicit symmetry breaking potentials $V(X)$ and $U(Y)$,
\bealg
\frac{R^3}{z^3}\rho_X^2(\partial_z\phi_X - q_A A_z)|_{\UV} &=& \frac{R^4}{z^4}\frac{\partial V}{\partial \phi_X}\bigg|_{\UV},\\
\quad \frac{R^3}{z^3}\rho_Y^2(\partial_z \phi_Y - q_B B_z)|_{\UV} &=& \frac{R^4}{z^4}\frac{\partial U}{\partial \phi_Y}\bigg|_{\UV}\,.
\eealg
Similarly, in the IR the variation of the action requires the boundary flux associated with the free linear combination of phases to vanish:
\beq
\left. \frac{R^3}{z^3} \left[ \kappa_A g_\theta^2 (\partial_z \theta - \kappa_A A_z) - q_A \rho_X^2 (\partial_z \phi_X - q_A A_z) \right] \right|_{\IR} = 0\,.
\eeq
The boundary condition on the PQ-brane is simply that the gauge-invariant CP-odd angle $\Omega$ remains continuous as one crosses the PQ brane. Since the $Y$ field terminates on the PQ brane, this boundary condition can be converted to the relation
 \beq 
\phi_Y|_{\PQ} = 0\ ,
\eeq 
which will ensure the smooth transition of the physical angle between around the PQ brane.

\section{The axion and $\eta'$ modes and the strong CP problem}
\label{sec:zeromodes}
The key to understanding the basic properties of this model is to identify the modes corresponding to the axion and the $\eta'$ of QCD. Those are the modes which in the absence of the anomalies and explicit UV breaking would be exactly massless, as they would be genuine Goldstone modes. Hence one should first identify the zero modes of the theory in the limit $\kappa_{A,B}\to 0, V(X), U(Y)\to 0$. These $\tilde{\eta}'$ and $\tilde{a}$ modes are given by (see App.~\ref{app:zeromode})
\begin{eqnarray}
       A_z (x,z) &=& A_z^{\eta'}(z) \, \tilde{\eta}'(x), \ \ 
    \phi_X ( x,z) = \phi^{\eta'}_X(z) \, \tilde{\eta}'(x),  \nonumber \\
    B_z (x,z) &=& B_z^{a}(z) \, \tilde{a}(x), \ \ 
    \phi_Y (x,z) = \phi_Y^{a}(z) \, \tilde{a}(x) 
\end{eqnarray}
with the wave functions explicitly given in \Eqs{eq:Azbeseel}{eq:phiXbeseel}.

Turning back on the anomalies $\kappa_{A,B}$ will introduce a mass term for these zero modes, and generate a single mass for one combination of $\tilde{\eta}'$ and $\tilde{a}$, which will be identified with the physical $\eta'$ field. We can nicely isolate the resulting mass term by noting that the effect of turning on $\kappa_{A,B}$ is to generate a mixing between the scalar $\theta$ and the zero modes $\tilde{\eta}', \tilde{a}$ via the term 
\begin{equation}
(\partial_z \theta - \kappa_A A_z  -\kappa_B B_z )^2
\end{equation}
in the Lagrangian.  
Since we are considering only the zero modes, which have no 
$z$-dependent bulk energy, we can apply the zero mode constraints 
\beq \partial_z\phi_X - q_A A_z=0, \qquad \partial_z \phi_Y - q_B B_z=0,
\eeq 
to rewrite this expression as
\begin{equation}
\left( \partial_z \theta - \frac{\kappa_A}{q_A} \partial_z \phi_X  -\frac{\kappa_B}{q_B} \partial_z \phi_Y \right)^2\,.
\end{equation}
The effect of the zero mode mixing can be absorbed into the bulk $\theta$ field via the field redefinition $\theta (x,z) \to \theta (x,z)- \frac{\kappa_A}{q_A} \phi^{\eta'}_X(z) \tilde{\eta}'(x) -\frac{\kappa_B}{q_B}\phi_Y^a (z)\tilde{a}(x)$. The bulk anomaly term then simply becomes $(\partial_z \theta)^2$, the IR boundary condition reduces to $\theta|_{\IR}=0$, and $\theta$ is continuous across the PQ brane. The only remaining effect of the zero mode mixing is then through the UV boundary value, 
\beq
\theta|_{\UV} \rightarrow \theta_{\QCD} - \frac{\kappa_A}{q_A}\phi_X|_{\UV} - \frac{\kappa_B}{q_B}\phi_Y|_{\UV}\,,
\eeq
so the problem reduces to pure Yang-Mills with a shifted $\theta$-angle. 

As shown in the Appendix~\ref{app:zeromode} the shift corresponds exactly to the physical zero modes normalized by their respective decay constants; hence, the shift in $\theta_{\QCD}$ will be given by\footnote{
The definition of $f_a$ 
is chosen to be consistent with that of $f_\eta'$; therefore it may differ from other conventions.}
\begin{equation}
\theta|_{\UV} \rightarrow \theta_{\QCD} -  \frac{\sqrt{q_A N_f} }{f_{\eta'}} \tilde{\eta}'(x) -  \frac{\sqrt{q_B N_{\PQ}}}{f_a}\tilde{a}(x)\,.
\end{equation}
where we used the relation $N_f = \kappa_A / q_A$ and $N_{\PQ} = \kappa_B / q_B$. We refer the reader to the Appendix for the detailed QCD matching of these bulk parameters. 

Since the vacuum energy of pure Yang-Mills is given by $\frac{1}{2}\chi_{\YM} \theta_{\QCD}^2$, we find that in this holographic model we indeed obtain the vacuum energy to be 
\beq
\frac{1}{2}\chi_{\YM} \left( \theta_{\QCD} -  \frac{\sqrt{q_A N_f} }{f_{\eta'}} \tilde{\eta}' -  \frac{\sqrt{q_B N_{\PQ}}}{f_a}\tilde{a} \right)^2\,.
\eeq
Here $\chi_{\YM}$ is the topological susceptibility of the pure Yang-Mills theory without any flavors, which can easily be obtained also in the holographic picture~\cite{Csaki:2026arv, Mishra:2026lvq}.  

We can see that as expected, the anomaly will generate a mass to one combination of the two zero modes, which we can identify with the $\eta'$ of QCD:
\begin{equation}
\eta' = \cos \alpha \, \tilde{\eta}'+\sin\alpha \, \tilde{a}, \ \ a= -\sin\alpha\,  \tilde{\eta}+\cos\alpha \, \tilde{a}\,,
\end{equation}
where the mixing angle is $\tan \alpha = \sqrt{\frac{q_B}{q_A}\frac{N_{\PQ}}{N_f}} f_{\eta'}/f_a \ll 1 $. The orthogonal state is identified with the physical axion $a$, which at this point remains exactly massless.

With this field rotation the contribution of the anomaly term will simply be 
\begin{equation}
    \frac{1}{2} \chi_{\YM} \left( \theta_{\QCD}-\frac{\sqrt{2N_f}}{\tilde{f}_{\eta'}}\eta' \right)^2\,,
\end{equation}
with $\tilde{f}_{\eta'} = f_a f_{\eta'}/\sqrt{f_a^2+ \frac{q_B}{q_A} \frac{N_{\PQ}}{N_f}f_{\eta'}^2}\approx f_{\eta'}$.

Turning on explicit axial symmetry breaking ($m_q \neq 0$) introduces a localized UV boundary potential $V = -\sqrt{2}|m_q| R^{-\frac{3}{2}} \rho_X|_{\UV} \cos (\phi_X|_{\UV} - \theta_q)$ with $\theta_q \equiv \arg m_q$, yielding the explicit mass contribution to the 4D effective potential:
\begin{equation}
-|m_q|v_X z_{\IR}^{-2} \cos\left( \frac{2}{\sqrt{2N_f} f_{\eta^{\prime}}} \tilde{\eta}' - \theta_q \right)\,.
\end{equation}

Rotating the holographic basis states into the physical mass eigenstates $\eta'$ and $a$ allows us to rewrite $\tilde{\eta}'$ in terms of these modes. At energies below the $\eta'$ mass, integrating out the heavy $\eta'$ field captures the full $\theta$ dependence of the theory and naturally introduces the invariant CP-violating parameter,
\beq
\bar{\theta} \equiv \theta_{\QCD} - \frac{\kappa_{A}}{q_{A}} \theta_q.
\eeq
Substituting the rotated fields and integrating out the $\eta^{\prime}$ yields a final effective potential in terms of $\bar{\theta}$ and the axion $a$ only:
\beq 
V_{\text{eff}}(a) \simeq -\chi_{\QCD} \cos\left( \bar{\theta} - \frac{\sqrt{q_B N_{\PQ}}}{f_{a}} a \right)\,.
\eeq
Here, the screened topological susceptibility is $\chi_{\rm QCD} = \left( {\chi_{\YM}}^{-1} + {\chi_{q}}^{-1} \right)^{-1}$, where $\chi_{q} \propto |m_q|v_X z_{\IR}^{-2}$ is the explicit chiral symmetry breaking contribution obtained from the quark mass term. Minimizing this potential directly exhibits the adjustment of the axion VEV,
\beq
\langle a \rangle = \frac{f_{a}}{\sqrt{q_B N_{\PQ}}} \bar{\theta}\,,
\eeq
which dynamically cancels the CP-violating phase. From this effective potential, we directly obtain the physical axion mass:
\beq
m_a^2 f_a^2 = q_B N_{\PQ} \chi_{\QCD} + (\text{contribution from } U)\,.
\eeq
The first term reproduces the standard QCD axion mass driven by the screened topological susceptibility ($\chi_{\QCD}$). The second term encapsulates the explicit breaking effects sourced by the boundary potential $U(Y)$ describing the explicit PQ breaking. Protecting the axion mass -- and thereby solving the axion quality problem -- requires this second explicit breaking term to be exponentially suppressed. We will show below that this condition  is naturally achieved in the limit of large dimensions $\Delta_Y$ of the $Y$ field, essentially rendering the axion itself composite as well. 

\section{The axion quality problem and the composition of the axion}
The robustness of the axion solution depends on the Peccei-Quinn symmetry being exact. However, global symmetries are generally expected to be explicitly broken by UV physics, such as quantum gravity effects at the Planck scale. We parameterize this explicit breaking via a localized potential on the UV boundary:
\beq
U(Y) \supset -\frac{\epsilon_{\cancel{\PQ}}}{z_{\UV}^3} (R^{3/2} Y)^n + h.c.,
\eeq
where $\epsilon_{\cancel\PQ}$ characterizes the breaking strength with dimension $[\epsilon_{\cancel\PQ}]=1$, 
while $n$ is the dimension of the operator breaking the PQ symmetry (for example enforced by some $\mathbb{Z}_n$ discrete symmetry). Such potential will shift the minimum of the vacuum energy, inducing a non-zero CP-violating phase $\Delta\overline{\theta}$. 

From the general scalar profile $\rho_Y(z)$, we can see that the degree to which this boundary breaking ruins the axion adjustment mechanism depends on the explicit breaking source mode $s_Y$. This source is dynamically determined by the UV boundary condition evaluated modified by the presence of the breaking potential $U(Y)$. This modified UV boundary condition sets
\beq
s_Y \sim \epsilon_{\cancel{\PQ}} \b{ \frac{z_{\UV}}{z_{\PQ}} }^{\Delta_Y(n-1)}.
\eeq

The physical axion state is a linear combination of the bulk gauge field $B_z$ and the PQ scalar phase $a$. Because the gauge field is protected by the exact 5D bulk gauge symmetry, explicit breaking can only affect the axion through its scalar component. The resulting shift in the CP-violating vacuum angle, $\Delta\overline{\theta}$, is directly proportional to this explicit breaking mass contribution relative to the total mass squared:
\bealg
&\Delta\bar{\theta} \simeq \frac{1}{m_a^2 f_a^2} \left( \frac{q_B n^2\epsilon_{\cancel{\PQ}}}{N_{\PQ} z_{\UV}^3} \right) \left( v_Y z_{\PQ} \right)^n\left(\frac{z_{\UV}}{z_{\PQ}}\right)^{n\Delta_Y} \\
&\sim 10^{-10} \left( \frac{n}{N_\PQ}\right)^2\bigg( \frac{\epsilon_{\cancel{\PQ}}}{10^{12}\gev} \bigg)  \left( v_Y z_{\PQ} \right)^n \left( 10^{-7} \right)^{n\Delta_Y - 12}\,,
\eealg
where the numerical estimate assumes that $m_a^2 f_a^2 \simeq q_B N_{\PQ}\chi_{\QCD}$, $\chi_{\QCD} \approx (75.6 \mev)^4$, $z_{\UV}^{-1} = M_{pl}$, and $z_{\PQ}^{-1} = 10^{12} \gev$.
To successfully solve the strong CP problem, this shift must be bounded by $\Delta\overline{\theta} \le 10^{-10}$.

\begin{figure}[t!]
  \centering
  \includegraphics[width=0.9\columnwidth]{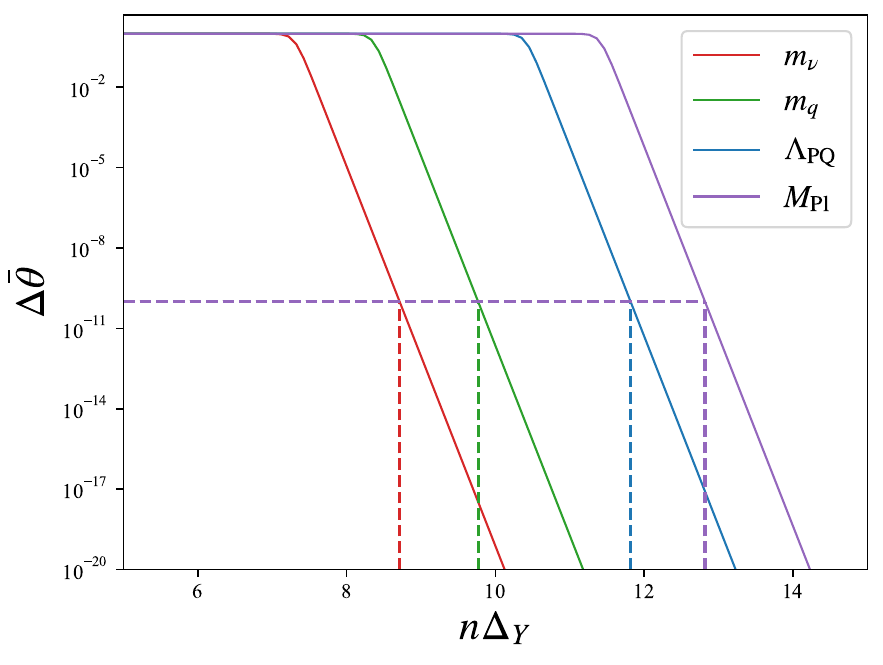}
  \caption{\protect\RaggedRight $\Delta\overline{\theta}$ against $n\Delta_Y$ for $ n = N_{\PQ}$, $v_Y z_{\PQ} = 1$ and different $\epsilon_{\cancel{\PQ}}$ values: $0.1 \ev$ ($m_\nu$), $3\mev$ ($m_q$), $10^{12}\gev$ ($\Lambda_{\PQ}$), and $1.22\times10^{19}\gev$ ($M_{pl}$). }
  \label{fig:rpq}
\end{figure}

One can see that the solution to the quality problem is strongly relying on choosing a large value for $n \Delta_Y \gsim 12$. This is illustrated in FIG.~\ref{fig:rpq}, where we show the induced shift $\Delta\overline{\theta}$ against the total suppression exponent $n\Delta_Y$ for various explicit breaking scales. One can see that satisfying the bound $\Delta\overline{\theta} \le 10^{-10}$ for small $\Delta_Y$ requires a very high order of discrete symmetry, while the large-$\Delta_Y$ limit can achieve this suppression more naturally. This choice of a very large $\Delta_Y$ implies that the axion itself should be strongly peaked on the PQ brane, and only a very small exponential tail left over on the UV brane. In the CFT language this is equivalent to requiring that the dimension of the operator describing the axion is very large, ie. the axion field itself is composite with the scale of compositeness being around $z_{\PQ}^{-1}$. Our results regarding the quality problem are very similar to those obtained in~\cite{Cox:2019rro} who examined generic warped extra dimensional axions with gluons either in the bulk or on the UV brane.

One additional interesting question is what the composition of the physical axion is in the limit when the quality problem is solved. The axion is partly contained in the bulk scalar $Y$ and partly in the $B_z$ gauge field. The relative composition can be defined as the ratio of their contributions of the normalization of the 4D kinetic term of the axion:
\begin{equation}
N_\text{axion} = N_B + N_a = \int_{z_{\UV}}^{z_{\PQ}} dz \frac{R}{e_B^2 z} B_z^2 + \int_{z_{\UV}}^{z_{\PQ}} dz \frac{R^3}{z^3} \rho_Y^2 \phi_Y^2\,.
\end{equation}
The relative gauge fraction, defined as $r_B \equiv N_B / (N_B + N_a)$, depends then on the dimensionless  parameter  
\beq
w_{\PQ} = \frac{\sqrt{2} q_B e_B v_Y z_{\PQ}}{\Delta_Y \sqrt{2R}}\,.
\eeq

In the limit when $w_{\PQ} \gg 1$ the $r_B$ ratio is dominated by the gauge contribution 
 \begin{equation}
r_{B} \approx \frac{\Delta_Y - 1}{\Delta_Y}
\end{equation}
implying that the axion is more than 90\% $B_z$ and only a smaller fraction arises from the scalar phase. Based on the QCD analogy one would expect $w_{\PQ}\gg 1 $ to be the relevant limit. 
Hence we conclude that high quality holographic axions are predominantly $B_5$ ``stringy axions". 
If however one chooses paramters for the axion sector where $w_{\PQ}<1 $ then this parameter will suppress the $B_z$ gauge contribution to the physical axion:  
\begin{equation}
r_B \approx \frac{\Delta_Y(\Delta_Y-1)}{2(2\Delta_Y-1)} w_{\PQ}^2 = \frac{\Delta_Y-1}{\Delta_Y(2\Delta_Y-1)} \frac{q_B^2 e_B^2 v_Y^2 z_{\PQ}^2}{4R}\,.
\end{equation}

\section{The Holographic PQ mechanism}
In the preceding sections, we addressed the strong CP problem and derived the axion mass by analyzing the zero-mode fluctuation spectrum in the 4D effective Lagrangian. However, a distinctive feature of this 5D construction is that the dynamical relaxation of the vacuum energy admits a purely geometric description. In this section, we demonstrate how the strong CP solution emerges in a manifestly holographic manner directly from the bulk background configurations.

To determine the vacuum structure and demonstrate the dynamical relaxation of the vacuum energy, we analyze the static ($k^2=0$) equations of motion. It is convenient to express these in terms of the conserved fluxes (canonical momenta) associated with the angular fields:
\bea
\Pi_\theta &=& \frac{R^3}{z^3} g_\theta^2 (\pd_z\theta - \kappa_A \, A_z - \kappa_B \, B_z), \\
\Pi_{\phi_X} &=& \frac{R^3}{z^3} \rho_X^2 (\pd_z\phi_X - q_A \, A_z),\\
\Pi_{\phi_Y} &=& \frac{R^3}{z^3} \rho_Y^2 (\pd_z\phi_Y - q_B \, B_z)\,.
\eea
Since the angles only appear through derivatives in the bulk action, their equations of motion are simply conservation laws:
\beq
\pd_z \Pi_\theta = 0, \qquad \pd_z  \Pi_{\phi_X} = 0, \qquad \pd_z \Pi_{\phi_Y} = 0\,.
\label{eq:pirelations}
\eeq
Furthermore, variations of the action with respect to the radial gauge fields $A_z$ and $B_z$ impose algebraic constraints that rigidly link these constant momenta across the bulk:
\beq
\kappa_A \, \Pi_\theta + q_A\, \Pi_{\phi_X} = 0, \qquad \kappa_B \, \Pi_\theta + q_B\, \Pi_{\phi_Y} = 0\,.
\label{eq:AzBzEOM}
\eeq

To evaluate the vacuum energy, we examine the bulk evolution of the gauge-invariant CP-odd angle $\Omega(z)$ definied in~(\ref{Omega}). Differentiating $\Omega(z)$ with respect to $z$ and substituting the momentum constraints to eliminate the gauge fields, one finds that its bulk profile is fully determined by the background value of $\Pi_\theta$:
\beq
\pd_z \Omega(z)= \frac{z^3}{R^3}\,\Pi_\theta
\left(\frac{1}{g_\theta^2}+\frac{\kappa_A^2}{q_A^2}\frac{1}{\rho_X^2}+\frac{\kappa_B^2}{q_B^2}\frac{1}{\rho_Y^2}\right)\,.\label{Omegaprime}
\eeq
This expression is almost identical to the bulk vacuum energy density, which can then be obtained by integrating it from the UV  to the IR. Using the boundary conditions Recalling the boundary conditions we find
\bea
\mathcal E_{\rm bulk}
&=&\int_{z_\UV}^{z_\IR} \dd z
\frac{z^3}{2R^3}
\left(
\frac{\Pi_\theta^2}{g_\theta^2}
+\frac{\Pi_{\phi_X}^2}{\rho_X^2}
+\frac{\Pi_{\phi_Y}^2}{\rho_Y^2}
\right) \nonumber \\
&=&\frac12\int \dd z \, \Pi_\theta \, \Omega'(z)
=\frac12\Pi_\theta\big(\Omega_\IR-\Omega_\UV\big) \,.
\label{eq:Ebulk}
\eea
As expected, the vacuum energy is a pure boundary term. 
To solve the strong CP problem, the axion field must relax the vacuum to a state where the energy is minimized. In the  limit where explicit PQ breaking on the UV brane is negligible (setting $U = 0$), the UV boundary condition will force the axion momentum to vanish
\beq
\Pi_{\phi_Y}|_{\UV} = 0 \implies \Pi_{\phi_Y} = 0\,.
\eeq
Then the algebraic constraints (\ref{eq:AzBzEOM}) linking the bulk sectors propagate this condition to the rest of the system:
\beq
\Pi_{\phi_Y} = 0 \implies \Pi_\theta = 0 \implies \Pi_{\phi_X} = 0\,.
\eeq
Consequently, the UV potential breaking the axial symmetry (ie. the quark mass terms) is also driven to its minimum via the BC $\partial V / \partial \phi_X = \Pi_{\phi_X} = 0$). With the vacuum momentum $\Pi_\theta$ vanishing identically across the bulk, the topological contribution to the energy density disappears. This confirms that the axion perfectly screens the external $\theta$-source, resulting in a vanishing topological susceptibility.

Setting $\Pi_\theta = 0$ in the integrated physical angle relation ($\Omega_{\IR} - \Omega_{\UV}$), we find that the boundary value of the axion dynamically adjusts to
\beq
\phi_Y|_{\UV} = \frac{q_B}{\kappa_B} \left( \theta|_{\UV} - \frac{\kappa_A}{q_A}\phi_X|_{\UV} \right)\,,
\eeq
showing that the holographic axion successfully relaxes the effective $\theta$ angle to zero, dynamically restoring CP invariance.

\section{Conclusion}
 In this work, we presented a holographic construction of the QCD axion in a warped 5D space, 
showing that the 5D geometry provides a natural framework for understanding the origin of the QCD axion mass. 
In this holographic description, the axion mass emerges from the QCD anomaly through its mixing with the $\eta'$ meson, 
while UV effects associated with Peccei–Quinn symmetry breaking encode the impact of quantum gravity on the axion potential.

A key ingredient of the construction is the 5D scalar field $\theta$, which is holographically dual to the QCD topological operator 
$\mathrm{Tr}\, G\tilde{G}$. A bulk St\"uckelberg term 
$(\partial_M \theta - \kappa_A A_M - \kappa_B B_M)^2$
generates the mixing between $\theta$, the $\eta'$ mode, and the axion. 
This provides a geometric realization of the anomaly-induced mass for the combination of $\eta'$ and axion. 
After including the explicit chiral symmetry breaking from the UV potential $V(X)$, the effective 4D theory reproduces the 
standard QCD axion potential. Additional explicit breaking from the PQ potential $U(Y)$ captures the effects of UV physics 
responsible for the axion quality problem.

Our analysis shows that the quality problem can be resolved when the product of the operator power and 
the conformal dimension of the PQ-breaking operator satisfies
$n \Delta_Y \gtrsim 12$.
We also find that in the high-quality limit, the physical axion is predominantly contained in the bulk gauge field $B_z$, 
while the scalar phase of $Y$ contributes a subdominant component. 
This indicates that the holographic axion behaves essentially as a ``stringy axion" arising from a higher-dimensional gauge field.

The holographic framework developed here opens several interesting directions for future study. For example, 
introducing a black hole horizon in the warped background would allow one to investigate the finite-temperature behavior of the axion potential 
and its evolution across the QCD phase transition. It would also be interesting to study axion string configurations within 
the full 5D geometry, which may provide new insights into axion cosmology and topological defects in holographic QCD.

\section{Acknowledgments}
\begin{acknowledgments}
The authors thank Gregory Gabadadze, Liam McAllister, Lisa Randall, Pierre Sikivie, Raman Sundrum, Matt Strassler, Ofri Telem, Charles Thorn for helpful discussions and feedback.
CC and TY are supported in part by the NSF grant PHY-2309456. CC and EK are supported in part by grants No 2022713 and 2024091 from the US-Israel BSF. EK is also supported by grant No 2023711 from the US-Israel BSF. TY is also supported in part by the Samsung Science and Technology Foundation under Project Number SSTF-BA2201-06. WX is supported in part by the U.S. Department of Energy under grant DE-SC0022148 at the University of Florida.
Part of this research was performed at the Munich Institute for Astro-, Particle and BioPhysics (MIAPbP)  funded by the  DFG  grant EXC-2094 – 390783311, and in part at the Aspen Center for Physics, supported by the NSF grant PHY-2210452 and the  Simons Foundation grant 1161654. EK is grateful to Cornell University for its hospitality and support during a sabbatical visit.

\end{acknowledgments}

\appendix
\setcounter{secnumdepth}{2}

\section{QCD Matching}

\paragraph{{\bf Bulk Gauge Coupling ($g_\theta^2$)}}
The coupling for the $\theta$ field is determined by the two-point correlation function of the topological charge density in pure Yang-Mills theory~\cite{Katz:2007tf}. The perturbative result at large momentum $Q$ is $\chi_\text{top}(Q^2) \propto -N_c^2 \frac{\alpha_s^2}{32\pi^4} Q^4 \ln Q^2$. Calculating the holographic correlator from the bulk action $\int \frac{g_\theta^2}{2} (\partial \theta)^2$ and matching the coefficients yields
\beq
g_\theta^2 = \frac{N_c^2 \alpha_s^2}{4\pi^4 R^3}.
\eeq
Note however that this is a perturbative matching assuming $Q^2\gg 1/z_{IR}$, and a large non-perturbative running is expected to affect $g_\theta$. It's low energy value relevant for axion physics is essentially a free parameter in this model. 

\paragraph{{\bf Anomaly Coefficients ($\kappa_A, \kappa_B$)}}
The divergence of the singlet axial current sums the anomaly contributions from all $N_f$ flavors: $\partial_\mu J^\mu_5 = 2N_f \frac{\alpha_s}{8\pi} G\tilde{G}$. Under a chiral rotation by angle $\epsilon_A$, the path integral measure shifts the vacuum angle by $\theta_{\QCD} \to \theta_{\QCD} + q_A N_f \epsilon_A$. In the holographic dual, a bulk gauge transformation $\kappa_B$ shifts the Stueckelberg field by $\delta \theta = \kappa_A \, \epsilon_A$. Identifying the transformation properties implies the matching condition:
\beq
\kappa_A = q_A N_f.
\eeq
By an analogous argument for the Peccei-Quinn symmetry, we identify the PQ anomaly coefficient:
\beq
\kappa_B = q_B N_{PQ}.
\eeq

\paragraph{{\bf Gauge Couplings ($e_A^2, e_B^2$)}}
The coupling for the bulk axial gauge field is fixed by the two-point function of the flavor-singlet axial current $J_\mu^5$. The perturbative fermion loop calculation gives $\Pi_A(Q^2) = -\frac{N_c N_f}{12\pi^2} \ln Q^2$. Matching this to the holographic result derived from the bulk gauge kinetic term $-\frac{1}{4e_A^2} F^2$ determines the coupling:
\beq
e_A^2 = \frac{6\pi^2 R}{N_c N_f}.
\eeq
This coupling also determines the decay constant for the singlet meson
\beq
\sqrt{2N_f} f_{\eta'} = \lim_{z \to z_{\UV}} \frac{R}{e_A^2} \frac{A_z}{z}\,.
\label{eq:feta}
\eeq
Similarly, the PQ gauge coupling $e_B^2$ is determined by the analogous current correlator for the PQ sector, yielding the relation for the axion decay constant:
\beq
e_B^2 = \frac{6\pi^2 R}{N_{c, \PQ} N_{\PQ}}.
\eeq

\paragraph{{\bf $U(1)$ Charges ($q_A, q_B$)}}
The charge $q_A$ is determined by the transformation of the chiral condensate $\langle \bar{q} q \rangle$ under the axial symmetry. Since $\langle \bar{q} q \rangle \to e^{2i\epsilon_A} \langle \bar{q} q \rangle$, and the dual bulk scalar $X$ transforms as $X \to e^{iq_A\epsilon_A} X$, we identify the charge:
\beq
q_A = 2.
\eeq
Similarly, $q_B$ is the specific PQ charge of the scalar field $Y$ responsible for breaking the Peccei-Quinn symmetry.

\section{$R_\xi$ Gauge}
\label{app:RxiGauge}
To rigorously define the physical zero modes and justify the exact constraints utilized in the main text, we must eliminate the mixing between the longitudinal gauge components and the scalar phases. This is achieved by introducing a generalized $R_\xi$ gauge-fixing term to the bulk action. Focusing on the $U(1)_A$ chiral sector for clarity, the gauge-fixing action is given by
\bealg
S_\text{g.f.} =& -\frac{1}{2e_A^2 \xi} \int d^5x \sqrt{|g|} \left[ \frac{z^2}{R^2} \partial_\mu A^\mu - \xi \frac{z^3}{R^3} \partial_z \left(\frac{R}{z} A_z\right) \right. \\ 
&  \hskip 9em + e_A^2 \xi (g_\theta^2 \kappa_A \theta + \rho_X^2 q_A \phi_X) \bigg]^2\,.
\eealg
The terms multiplying $\xi$ define the gauge constraint $\mathcal{F}_A$
\beq
\frac{\mathcal{F}_A}{\xi} = -\frac{1}{e_A^2} \partial_z \left(\frac{R}{z} A_z\right) + \frac{R^3}{z^3} \rho_X^2 q_A \phi_X + \frac{R^3}{z^3} g_\theta^2 \kappa_A \theta\,.
\eeq
In $R_\xi$ gauge, the transverse and longitudinal components decouple. 

\section{Zero Mode Profiles}
\label{app:zeromode}
The bulk EOM's can be formulated in terms of the gauge invariant combinations 
\begin{eqnarray}
&\Theta &= \pd_z\theta -\kappa_A A_z -\kappa_B B_z \nonumber \\
&\Phi_X &= \pd_z\phi_X-q_A A_z \nonumber \\
&\Phi_Y& =\pd_z\phi_Y-q_B B_z\ .
\end{eqnarray}
The zero modes governing the chiral dyanmics have the property that the gauge invariant combinations defined above vanish identically $\Phi_X=\Phi_Y=0$. This will yield the constraint connection the gauge field and the phase for the zero modes:
\beq
\partial_z \phi_X = q_A A_z, \qquad \partial_z \phi_Y = q_B B_z\, .
\label{eq:zeromodeconstraint}
\eeq
In addition the gauge fixing term evaluated on the zero modes should be vanishing, otherwise it could not correspond to a physical mode. Our expectation is that in the limit of vanishing anomalies $\kappa_{A,B}\to 0$ and boundary terms $V,U\to 0$ there should be two exact zero modes in the system - the Goldstone bosons corresponding to the spontaneously broken $U(1)_A$ and $U(1)_{PQ}$. Here we will explicitly find the profiles of these two modes $\tilde{\eta}', \tilde{a}$, which are used in the main text. 
These two modes will be identified as the $\eta'$ and axion modes prior to picking up a mass and mixing due to the anomalies and the explicit breaking terms.  For simplicity we will focus on the $\tilde{\eta}'$ mode, the $\tilde{a}$ mode is obtained analogously. The vanishing of the gauge fixing constraint $\mathcal{F}_A = 0$ in the absence of anomalies gives 
\beq
-\frac{1}{e_A^2} \partial_z \left(\frac{R}{z} A_z\right) + \frac{R^3}{z^3} \rho_X^2 q_A \phi_X = 0\,.
\label{eq:FAAzphiX}
\eeq
Incorporating a generalized background chiral condensate profile $\rho_X(z) \approx \rho_{X,0} z^\Delta$, we can solve this constraint for the Goldstone phase $\phi_X$:
\beq
\phi_X(z) = \frac{1}{R^3 \rho_{X,0}^2 q_A e_A^2} z^{3-2\Delta} \partial_z \left(\frac{R}{z} A_z\right)\,.
\eeq
Together with the zero mode constraint $\partial_z \phi_X = q_A A_z$ we find the equation for the zero mode wave function
\beq
\partial_z \left[ z^{3-2\Delta} \partial_z \left( \frac{A_z}{z} \right) \right] - \beta^2 A_z = 0\,,
\eeq
where $\beta^2 = q_A^2 e_A^2 R^2 \rho_{X,0}^2$. Defining a rescaled wave function $f(z)$ such that $A_z(z) = \frac{z}{R}f(z)$, the equation turns into the modified Bessel differential equation $z^2 f^{\prime\prime}(z) + (3-2\Delta)z f^\prime(z) - \beta^2 z^{2\Delta} f(z) = 0$.

Thus the explicit zero-mode profile for the $A_z$ gauge field is  
\beq
A_z^{\eta'}(z) = \mathcal{N} z^\Delta \left[ K_\nu(w_{\IR}) I_\nu(w(z)) - I_\nu(w_{\IR}) K_\nu(w(z)) \right]\,,
\label{eq:Azbeseel}
\eeq
where the order is $\nu = \frac{\Delta-1}{\Delta}$ and the argument is $w(z) = \frac{\beta}{\Delta} z^\Delta$.
The component of the zero mode living inside the scalar $\phi$ then follows from (\ref{eq:zeromodeconstraint}):
\beq
\phi^{\eta'}_X(z) = \mathcal{N} \frac{q_A}{\beta} z \left[ K_\nu(w_{\IR}) I_{\nu-1}(w(z)) + I_\nu(w_{\IR}) K_{\nu-1}(w(z)) \right]\,,
\label{eq:phiXbeseel}
\eeq
One integration constants has been determined via the IR Dirichlet boundary condition $A_z(z_{\IR}) = 0$. 
The remaining normalization constant ${\cal N}$ is obtained by requiring that the 4D kinetic term of the zero mode is properly normalized, leading to 
\begin{equation}
{\cal N} = \left[ - \lim_{z \to z_{\UV}} \frac{R}{q_A e_A^2 z} \left(\frac{A_z^{\eta'}(z)}{\mathcal{N}}\right) \left(\frac{\phi_X^{\eta'}(z)}{\mathcal{N}}\right) \right]^{-1/2}\,.
\end{equation}
Note that ${\cal N}$ is related to the decay constant 
$f_{\eta'}$ from \eqref{eq:feta} as 
\begin{equation}
\mathcal{N} = \frac{e_A^2 \sqrt{2N_f} f_{\eta^{\prime}}}{R \, z_{UV}^{\Delta-1} \Big[ K_\nu(w_{IR}) I_\nu(w_{UV}) - I_\nu(w_{IR}) K_\nu(w_{UV}) \Big]} \,.
\end{equation}
In the limit $z_{\UV} \to 0$, the $I_\nu(w_{\UV})$ term mathematically vanishes. Substituting the small-argument expansion $K_\nu(w_{\UV}) \simeq \frac{\Gamma(\nu)}{2} \left(\frac{w_{\UV}}{2}\right)^{-\nu}$ reveals a perfect cancellation of the remaining $z_{\UV}$ dependence, yielding
\begin{equation}
\mathcal{N} = - \frac{e_A^2 \sqrt{2N_f} f_{\eta'}}{R \, I_\nu(w_{\IR}) \frac{\Gamma(\nu)}{2} \left(\frac{2\Delta}{\beta}\right)^\nu} \,.
\end{equation}
Hence the full expression of the $\eta'$ mode is
\begin{equation}
    A_z (x,z) = A_z^{\eta'}(z) \tilde{\eta}'(x), \ \ 
    \phi_X (x,z) = \phi^{\eta'}_X(z) \tilde{\eta}'(x)
\end{equation}
with the 4D mode satisfying $\Box_4 \tilde{\eta}' = 0$
Thus we have identified the zero mode for arbitrary $\Delta$ (setting the shape of the bulk scalar VEV and corresponding to the dimension of the quark condensate). An identical derivation holds for the Peccei-Quinn sector gauge field $B_z$ and the axion phase $\phi_Y$, leading to a zero mode 
\begin{equation}
    B_z (x,z) = B_z^{a}(z) \, \tilde{a}(x), \ \ 
    \phi_Y (x,z) = \phi_Y^{a}(z) \, \tilde{a}(x)
\end{equation}
with $B_z^a$ and $\phi_Y^a$ as in (\ref{eq:FAAzphiX}-\ref{eq:phiXbeseel})
with the replacements $\beta^2\to q_B^2e_B^2 R^2 \rho_{Y,0}^2$, $z_{\IR}\to z_{\PQ}$, $\Delta_X\to \Delta_Y$.

\bibliography{bibtex}

\onecolumngrid

\end{document}